\documentclass{article}

\usepackage{amsmath}
\usepackage{listings}
\lstdefinelanguage{scala}{
  morekeywords={abstract,case,catch,class,def,%
    do,else,extends,false,final,finally,%
    for,if,implicit,import,match,mixin,%
    new,null,object,override,package,%
    private,protected,requires,return,sealed,%
    super,this,throw,trait,true,try,%
    type,val,var,while,with,yield},
  otherkeywords={=>,<-,<\%,<:,>:,\#,@},
  sensitive=true,
  morecomment=[l]{//},
  morecomment=[n]{/*}{*/},
  morestring=[b]",
  morestring=[b]',
  morestring=[b]"""
}

\usepackage{hyperref}

\begin{document}

\title{annotated imports }

\author{Ruslan Shevchenko}
           {ruslan@shevchenko.kiev.ua}

\maketitle

\begin{abstract}

  Presented simple extensions to scala language related to import statements:  
 exported imports, which provide ability to reuse sequence of import clauses in composable  
 form and default rewriters, which provide mechanism for pluggable macro-based AST  
 transformation of overall compilation unit, activated by import of library object.
 
  Using these facilities not only allows more compact code, it prevents application programmer from producing certain type of errors too and allows to implement
local language extension as libraries on top of standard compiler.

 Part of discussed extensions is submitted to scala language committee as pre-sip \cite{ai-presip} and can be used as first step for refining imports semantics in the future version of scala language.

\end{abstract}




\section{Composability}

  Currenty import statementes are not 'first-class-sitizens' of scala-language, the main unsupported property is  composability -  i.e. now it is impossible to unite sequence of import statements into one reusable construction.

  Absence of such features is a source not only for code bloat 
  \footnote{actually not so big, but can be not so small: import statements usually occupied from 2 (scala-compiler) to 15 (play-mustasche plugin) percents of code size}, but also can be source of particular type of errors, which is hard to detect in compile time. 
  Sequence of imports in program determinate order of default implicit resolutions, so absence of ability to squeeze sequence of imports into one give us space for gotchas like next:

File A:
\begin{lstlisting}
import com.mongodb.casbah.Imports._
import com.novus.salat._
// customized database context
import  com.mycompany.salatcontext._
   …..............
 
\end{lstlisting}
  
File B:

\begin{lstlisting}
import com.mongodb.casbah.Imports._
import com.novus.salat._
// standard database context
import com.novus.salat.global._
   …..............
 
\end{lstlisting}
  
 Here in file A we use custom database context, in B -- standard. Compiler can't know that  we want one version of salat context across all project, therefore our project, which include both A and B, will be compiled fine but will fail in runtime with cryptic error message.

The proposed feature, called  annotated import, is to  allow in grammar to put static annotations before import statement and introduce special annotation @exported for  import specifications  (proposed syntax:  @exported import ) inside object scope,  which  export content of specification to all contexts which import this scope with next semantics: 
\begin{itemize}
 \item @exported imports must be situated inside templates (i.e. classes and objects)
 \item if template imported by wildcard, then search in this template object includes search in all @export-annotated imports in this template object.
 \item loops in @exported imports are allowed with usual rules for preventing infinite recursion.
\end{itemize}

 For previously described example, natural way will be to define all imports in one package object

\begin{lstlisting}
package com.mycompany
package object salat {
@exported import com.mongodb.casbah.Imports._
@exported import com.novus.salat._
@exported import com.mycompany.salat.context._
}
\end{lstlisting}

and then import this package object in files A and B.

Files A, B:
\begin{lstlisting}
import com.mycompany.salat._
   …..............
 
\end{lstlisting}

In general, we think that exported imports must be used not during export of library interfaces by library authors, but during import of ones from application layer. 

From architecture point of view, we expect that non-trivial applications will
 contains layer for imports of external dependencies and internal application code
  will use external facilities only via this layer.

\subsection{Passing exported imports by inheritance}
  
  Visibility of exported imports by inheritance will help in reverse situation: i.e. when library author want to reduce amount of work, needed for using his library.  In such
 case author of class library can provide all necessary environment for implementors of inherited classes and objects. This is particularly interesting for using in frameworks (such as Play), which assume that users must implement own version of interfaces,  predefined in this framework. 
  
\begin{lstlisting}
trait Controller
{
 @exported import play.api._
 @exported import play.api.mvc._

 
}
\end{lstlisting}

and then in clients:
\begin{lstlisting}
object MyContoller extends play.api.mvc.Controller
{

 // use exported api here. 
 ... 
 
}
\end{lstlisting}

Note, that quite many existing frameworks now emulate export by inheritance via defining set of common functionality (such as definitions of functions and case classes) in traits for possibility to use one in objects, inherited from this trait. This is exactly case for implicit passing of exported import. The second will not force programmer to distort object model and will not generate boilerplate bytecode for bridge methods in each trait incarnation.

\subsection{Compiler API changes:}

 Implementation require changing one rule in grammar: from
\begin{lstlisting}
TemplateStat ::=   Import
                 | {Annotation} {Modifier} Def
                 | {Annotation} {Modifier} Dcl
                 | Expr
\end{lstlisting}

to
\begin{lstlisting}
TemplateStat ::=    {Annotation}  Import
                             | {Annotation} {Modifier} Def
                             | {Annotation} {Modifier} Dcl
                             | Expr


\end{lstlisting}

 altering symbol representation and implementing keeping of exported imports into pickled signature.

Changes in compiler infrastructure where relative small: each symbol already contains annotations, tree representation can be described as combination of already existent tree elements, 
\begin{lstlisting}
Annotated(Annotation,ImportTerm)
\end{lstlisting}
 which is erased during typer phase.

\section{ Importing language feature }

Next simple extension: library-level definition of language dialects.  Current compiler supports specifying of language features (from predefined set, hardcoded in compiler),  through process of implicit resolution, described in SIP-18 \cite{SIP-18}. From other side,
we have macros reflection API, which allows us to describe transformations of AST tree inside compilers.  And near any language extension, compatible with existent language grammar, can be described in terms of such AST transformations.  So, it is possible to implement more dynamic configuration of compile-time behavior using the same process of implicit resolution.

Let's look at next example: 

\begin{lstlisting}
package copyfile

import java.io._
import go.defer._

object Main
{


  ...............

  def copy(inf: File, outf: File): Long =
  {
    val in = new FileInputStream(inf)
    defer{ in.close() }
    val out = new FileOutputStream(outf);
    defer{ out.close() }
    out.getChannel() transferFrom(in.getChannel(), 0, Long.MaxValue)
  }
  
}

\end{lstlisting}
 
 As you see, this code block use 'defer'  keyword in the same way, as it used in Go \cite{go} programming language: statement inside defer will be executed when execution flow 
 will leave method scope.
 
 Also note, that interpretation of defer as keyword is specified by importing 
 $'go.defer.\_'$  namespace.

 How this works -- AST transformations are performed using macro reflection api, defined as implicit object in 'go.defer'.

\begin{lstlisting}
  implicit object GoDefaultRewriter extends DefaultRewriter
  {

    override def transformAImpl(c:AnnotationContext): c.Tree =
       .............
  }
\end{lstlisting}

   We implemented extra-simple compiler plugin \cite{scala-language-import} on top of macro-paradise\cite{macro-paradise} which just adds static macroannotation to all classes and objects in compilation unit.  The work of this annotation is to find implicit rewriter in current scope, then instantiate and call one.
   
   Combination of two language extensions can be handled with help of next construction:
   
\begin{lstlisting}
package AwithB {
  @exported import A.{rewriter=>_,_}
  @exported import B.{rewriter=>_,_}
  implicit object rewriter extends DefaultRewriter
  {
    override def transformAImpl(c:AnnotationContext) =
       A.rewriter.transformAImpl(B.rewriter.transformAImol(c))
  }
}
\end{lstlisting}

  May be there is a sense to create something like combinators algebra for language extensions with rules of automatic combination for simple cases.
  
  Using such mechanism in language core allows to build extensions to scala language which does not extend original language grammar, such as scala-virtualized\cite{scala-virtualized}, to be implemented as macro libraries on top of standard compiler.

\subsection{Conclusion} 

 So, we have shown that semantics of import statements can be improved with help of relative-simple mechanisms: exported imports allows reuse sequence of imports and can be 
 helpful in situation when we need to configure compile-time context be the same across project;  ability to specify implicit rewriting rules allows library-based language extensions. 
 
 Future directions - enrich set of possible import annotations, particularly interesting points can be: calling external tools by compiler;  framework for combination of language extensions - can we define some generic rules for merging few AST transformation into one. Note, that this problem can be generalized to general rules of resolving ambiguous implicit-s:  i.e. for some type $T$ define $ImplicitltyConflictResolver[T]$ which
 will provide strategy for choosing one instance of implicit variable across set of resolved, possible using compile-time accessible properties of resolved instances.
 
 Yet one interesting research theme: think, how to make syntax representation more flat and move part of parser work (forming language constructions) into potentially extensible space of AST rewriting.


%
%
%
%

\bibliographystyle{abbrvnat}


\end{document}